# Social Event Detection with Interaction Graph Modeling

Yanxiang Wang
Australian National University
tauruswyx@gmail.com

Hari Sundaram
Arizona State University
hari.sundaram@asu.edu

Lexing Xie
Australian National University, NICTA
lexing.xie@anu.edu.au

## ABSTRACT

This paper focuses on detecting social, physical-world events from photos posted on social media sites. The problem is important: cheap media capture devices have significantly increased the number of photos shared on these sites. The main contribution of this paper is to incorporate online social interaction features in the detection of physical events. We believe that online social interaction reflect important signals among the participants on the "social affinity" of two photos, thereby helping event detection. We compute social affinity via a random-walk on a social interaction graph to determine similarity between two photos on the graph. We train a support vector machine classifier to combine the social affinity between photos and photo-centric metadata including time, location, tags and description. Incremental clustering is then used to group photos to event clusters. We have very good results on two large scale real-world datasets: Upcoming and MediaEval. We show an improvement between 0.06–0.10 in F1 on these datasets.

## 1. INTRODUCTION

The goal of this paper is to detect social, physical-world events from photos posted on social media sites including Flickr. The problem is important because the advent of cheap photo-capture devices, including mobile phones, has made it easy for people to capture and share media from physical events, such as concerts, festivals, and sports matches. This has resulted in a large number of photos being shared on social media sites. The large number of photos from such events makes it imperative that we are able to identify and organize event related photos.

Event detection on social media platforms has attracted considerable recent interest. Firan et al. [3] solves event detection as a supervised classification problem, where classifiers are built from the metadata of each event. Becker et. al. [1] poses this as an unsupervised clustering problem on social photo streams. This approach exploits a number of features including time, location and textual information. A classifier is trained to learn the similarity metric between photo pairs, and then photos are clustered into different event groups. Two other groups [4, 7] uses visual analysis to find photos to illustrate and group events, community detection [4] is further introduced to improve performance. In another recent work, Becker et al. [2] designed a two-step approach to first cluster the input twitter stream and then perform event versus non-event classification on the clusters. In all current approaches, photo content, tags, times and locations are used as the main features to cluster event photos. We note, however, that *social interactions* did not play a role in any of the prior approaches.

The main contribution of this paper is to incorporate social features in the detection of physical-events. Real-world events are intrinsically social—they bring people together to participate in a shared experience. Therefore, it is natural to expect that co-participation in the physical event will motivate people who attended the event to interact online with shared media associated with that event. We argue that the social interaction reflect important and explicit signals on part of the participants on the "social affinity" of two photos, thereby helping in event detection. The social affinity between photos, in addition to photo metadata including time, location, tags and description are used to detect events.

To compute social affinity between two photos, we proceed as follows. First we construct a directed, social interaction graph based on the relationship amongst users, tags and photos. Then, we use a random-walk model to determine similarity between two photos on the graph. This formulation appears naturally when we consider how users interact around shared photos with a social media site . Then, we train a SVM on features from the social interaction graph and base photo features that have been shown to work well on event detection [1]. We shall use an incremental clustering algorithm on the photos to find event clusters. We tested our algorithm on two large datasets—MediaEval and Upcoming—with very good results. In both cases, our approach improves upon the state of the art by 0.102 (Upcoming) and 0.059 (MediaEval) respectively.

The rest of the paper is organized as follows. In the next section, we introduce the idea of social affinity. Then, in Section 3, we discuss the photo-features used as a baseline method for social event detection. In Section 4, we present our incremental clustering algorithm to detect events. In Section 5, we present our experimental results. Finally, we conclude with our summary and conclusions.

## 2. SOCIAL AFFINITY

In this section we shall develop the idea of social affinity—*that is, can we relate two photos based on the social interaction around them?* Next, we discuss how we incorporate the social dimension in our measure of photo similarity.

### 2.1 How to Measure Social Affinity?

The Flickr API[1] exposes several "social" features for each photo—that is, data about activities around each photo. We list them below:

[1] http://www.flickr.com/services/api/

- **Photo comment**: Each photo comment is associated with the specific user, and a specific time stamp.
- **Photo favorite**: We know the set of users that have marked a specific photo as a favorite.
- **Photo tag**: Each photo is typically associated with a set of tags that are assigned by the owner or others.
- **Related tags** obtained via the *flickr.tags.getRelated* API method. These tags—unlike a thesaurus expansion—represent tags commonly co-occur with each other. An association between the tag “newyork” and “brooklynbridge” is an example.
- **Owner contact list**: We know the identity of the owner of each photo, as well as the list of users who are part of her contact list.

Given two photos, can the number of users who interact with both photos be a useful measure of affinity? Intuitively, the higher the number of common users for two photos, the more likely that the two photos are related to the same event. The Jaccard set similarity (Sec 3) score on the two sets of users associated with each photo, is a common measure of set similarity. There are two challenges. First, we need to deemphasize a spam-like commenter, who comments on each photo—this person lacks the ability to discriminate. Second, general tags such as ‘2011’, ‘Nikon’are useless as they appear on nearly every photo. Both of these challenges indicate that we need to emphasize users as well as tags that are both discriminative.

We propose a random-walk model to determine similarity between two photos. Assume that a user is interested in browsing photos from one particular event $E$. She can start from a particular photo $p_i \in E$ and perform several activities. First, she can randomly click on a tag, and find other event photos. Second, she can click on any user who has commented on $p_i$ and examine their photos. After examining related photos, she may return to the original photo, to follow other users or tags. Notice that in the implementation, we manually filter out spam commenter and spam tags. The user will never click on the spam user or the non-discriminant tag more than once. And when they do, since each spam-user or non-discriminative tag is associated with many photos, it is unlikely that a user will go through these large set of associated photos. Next, we explain how we construct the social interaction graph.

## 2.2 The Social Interaction Graph

We represent the social interaction around photos using a undirected graph with three conceptual layers: photos, users and tags. Each unique user, each photo and each tag form the vertices of the graph. The edges between the nodes can be symmetric or asymmetric, and additionally, the edge semantics can vary. Figure 1 illustrates a sample interaction graph. The similarity measure between two photos, is the probability that a user will “walk” from one photo to the other.

We create symmetric links in three cases. First, we put undirected edges between a photo and a tag associated with it. Second, we connect tags returned by the Flickr API that provides related tags . Adding this link simulates the behavior that people tend to explore related tags when browsing photos. Third, we create symmetric edges between a photo and each user who interacts with the photo, i.e.: the user can own the photo, leave comments or mark the photo as a favorite.

The relationship between two users $a$ and $b$ can result in an asymmetric link between them. Both users need to acknowledge the other as a contact, for the link to be symmetric.

We use a Random Walk with Restart algorithm (RWR) [5] on the social interaction graph to determine the affinity between two photos. In [5], the authors use RWR algorithm to successfully caption photos. The RWR algorithm calculates the steady state probability from the starting node $v_s$ to another node $v_e$ in the graph. Consider the following scenario: a random walker starts from node $v_s$. In each step, the walker chooses randomly among the available edges, with one modification: for each step, there is a probability $c$ that she goes back to starting node $v_s$. The affinity of node $v_e$ to starting node $v_s$ is then the steady state probability $p_s(v_e)$ that our random walker will reach node $v_e$.

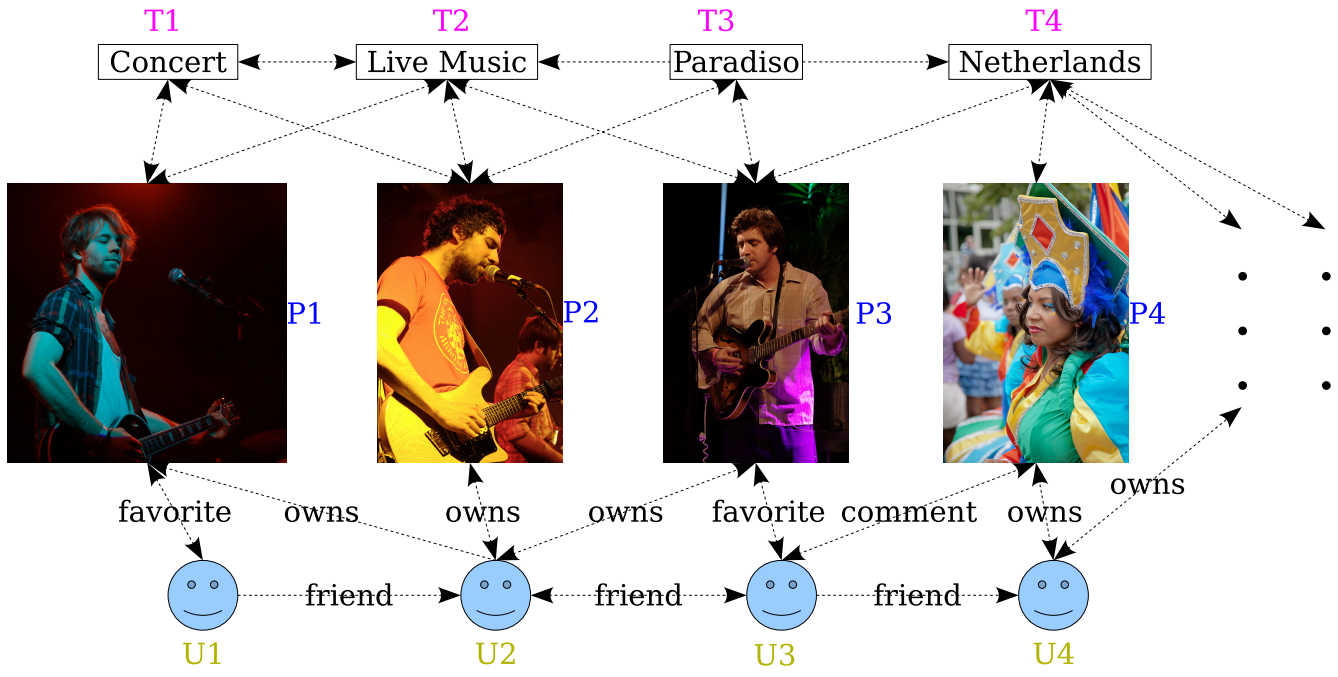

**Figure 1: Photo Graph Example. The edges can be symmetric (photo-user) as well as asymmetric (user-user). Photos courtesy of Flickr users Martijn vdS and Qsimple.**

The sequence of steps to compute affinity between a pair of photos $(v_s, v_e)$ are as follows:

1. Build the social interaction graph, resulting in the corresponding adjacency matrix $A$.
2. Create a restart vector $\vec{v}$ where all entries are set to zero except for the entry for $v_s$, which is set to 1.
3. Calculate the steady state vector $\vec{v_{ss}}$ using the following formula $\vec{v}_{ss} = c(I - (1-c)A)^{-1}\vec{v}$. The entry in the steady state vector corresponding to node $v_e$ will give us—after normalization—the social affinity score between photos $v_s$ and $v_e$.

The RWR algorithm creates a bias towards photos that are close to the start node. This is reasonable since we don’t expect a user to “wander off” to remote parts of the graph that are many hops away from the start node.

The size of the adjacency matrix $A$ (with photos, tags and users each in the tens of thousands), and the need to invert it, creates computational challenges for our system. In this work we use a sliding time window to limit the size of the adjacency matrix, and compute the social interaction graph and the corresponding matrix, per window. We use overlapping windows of duration two weeks, with one week of overlap between adjacent windows. The two-week duration is reasonable, since many social events of interest, such as sports games, music performances rarely exceed a day. Future work include address the selection of the interaction window, or suitable parallelization to scale up to larger matrixes.

# 3. PHOTO FEATURES

We can extract a set of features, including time, location, tags and description, for each photo. These features comprise the baseline and are derived from prior work [1] where these features yield the best performance for social event detection.

- **Time-stamp**: We represent time as the number of minutes elapsed since the beginning of the Unix epoch. We set the similarity of two photos to be 0, if two photos are taken more then a week apart. Otherwise, the similarity between two photos with taken-time $t_1$ and $t_2$ is computed as $\sigma_{time} = 1 - \frac{t_1 - t_2}{t_w}$, where $t_w$ as number of minutes in a week. We set the upper bound to be one week as most many events including music concerts, soccer games, last for less than a week.

- **Location**: The location similarity between two photos are computed using the Great Circle Distance [8]. We set the location similarity $\sigma_\ell$ to 0 if the $GCD$ value is greater than 50 miles (about 80 kilometers), otherwise $\sigma_\ell = 1 - \frac{GCD}{50}$. We set the threshold to be fifty miles, as this number is a reasonable estimate of the diameter of most cities[2]. In other words, if the two photo locations are in different cities, we set their location similarity value to be 0.
- **Tags**: For the tag similarity $\sigma_{tag}$, we use the Jaccard index to measure the similarity between the set of tags associated with each photo. Assume that photo A is tagged using tag set $t_A = \{t_1, \cdots, t_j\}$ and photo B is tagged using tag set $t_B = \{t_1, \cdots, t_k\}$. Then, the tag similarity $\sigma_{tag}(A, B)$ between two photo A and B, is $\frac{|t_A \cap t_B|}{|t_A \cup t_B|}$, where $|\cdot|$ is the cardinality of a set. We chose the Jaccard Index as it widely used to compute set similarity.
- **Text Description**: Each photo is associated with a title and text description. We obtain a term-frequency vector from the title and description of the photo after stemming and eliminating stop words. We compute the similarity value $\sigma_x(p_i, p_j)$ for two term-frequency vectors $\vec{v_i}$ and $\vec{v_j}$—derived from the corresponding photos $p_i, p_j$—using a cosine similarity measure.

These similarity values between photos $p_i, p_j$ are combined with the similarity measure from the social graph (see Section 2), to create the feature vector that is used in our clustering algorithm.

# 4. EVENT DETECTION

We learn first learn a photo similarity from multiple sources with Support Vector Machine (SVM) classifier, and then use an incremental clustering algorithm [1] to group photos into event clusters,

The similarity measure between photos is critical in determine the performance of the event detection algorithm. Given there are multiple features available for each photo from Flickr, we choose an SVM classifier to combine these features together and produce a single value similarity measure $\theta(p_i, p_j)$ between a pair of photos $p_i, p_j$.

On the training set, we first compile different features for classifier training. These include of similarity values from various photo and social features. While the similarity values of some photo features including time, photo-location are determined in a straightforward manner (Section 3), the social similarity is computed via a random-walk algorithm on the social graph as described in Section 2. The SVM classifier is trained on pairs of photos that either belong to the same event (positive), or are from different events (negative). We randomly sample training pairs from the dataset, and the ratio of the negative to positive pairs is kept below 5 to to avoid skewing the prior.

On the testing set, this SVM is applied on pairs of test photos, yielding a pair-wise similarity matrix. We then run a one-pass incremental clustering procedure to determine the event grouping. The algorithm incrementally assigns each photo $p_i$ into a cluster $C_k$ based on the similarity between this photo and existing clusters. To compute the photo–cluster similarity, we compute the average similarity between $p_i$ and each $p_j \in C_k$. Specifically, given a similarity function $\theta$, a threshold $\mu$ and the photo stream $p_1, p_2, \ldots, p_n$ We select the cluster that has the highest similarity value based on $\theta$ for each photo $p_i$ to be added. If no existing clusters whose similarity value to $p_i$ is greater than the threshold $\mu$, a new cluster is formed. We report results on a range of thresholds in the next section.

# 5. EXPERIMENTS

We evaluate our algorithm on two datasets: Upcoming [1] and the MediaEval [6] datasets. For both data sets, ground truth labels are available.

The Upcoming data set [1] contains around 270k photos crawled from Flickr and were tagged with an upcoming event (e.g. "upcoming:event=428084"). The dataset has 9,515 events and on average 28.42 photos per event, taken between January 1, 2006, and December 31, 2008. Within this set, 110,757 photos form 1,643 "social" events, i.e., events that are contributed by multiple participants. We use these photos to test our algorithms since they are more representative of a social event as more than one person participates and uploads photos. This set is split into 3 equal portions for training similarity metric, tuning cluster, and evaluating performance.

The MediaEval benchmark [6] consists of 73k photos taken in May, 2009 in the cities Paris, Rome, Barcelona and Amsterdam. The benchmark organizers manually generated labels for soccer games in Barcelona and concerts in two respective venues in Barcelona and Amsterdam for testing. There is no specifically provided training data, and event-specific photo tags and geo-tags have been removed or tampered to avoid making the task trivial.

We use a subset of Upcoming data as the training set for the SVM classifier. We train three variants of such classifiers (see Section 4). The first is features from the photo only, this is the same approach as [1] and consists of our baseline. A second baseline comes from naÃŕve social and photo features, here naÃŕve social features is the Jaccard similarly on the set of users that has owned or interacted with two photos (Section 2.1). Finally we combined social features from the interaction graph and photo features. The trained classifier is then tested on both the Upcoming and MediaEval datasets.

Not all the photos in the MediaEval data set—unlike the Upcoming dataset—correspond to real world events. We determine the correspondence to events in the following manner. After a clustering step on the MediaEval data-set, we perform a query-based retrieval method on the photo cluster metadata and report the performance on a range of different thresholds. Specifically, we collect information (via an API call) about soccer and concert events in the same cities and time-frame from Upcoming and Eventful websites. We rank the photo clusters based on the cosine distance between two bag-of-word vector: one compiled from each online event description, the other from words (title, description and tags) in the cluster.

We use the F1 metric on photo pairs to evaluate performance on both datasets. Figure 2 and Figure 3 show the respective evaluation results on the Upcoming and MediaEval datasets. We have plotted the F1 score while varying the incremental clustering threshold on the $x - axis$, this threshold helps determine whether to add a photo to an existing cluster, or to start a new cluster. In both cases, our approach based on a combination of the social interaction graph and the photo features, outperform the other two baseline approaches. On the Upcoming dataset, the maximum performance improvement over the baseline photo features is 0.102. For the MediaEval Dataset the maximum improvement is 0.059 over the baseline photo features.

Figure 4 shows example detection results for two upcoming events:

[2] The diameter of the Greater New York City is $\approx 20$ miles.

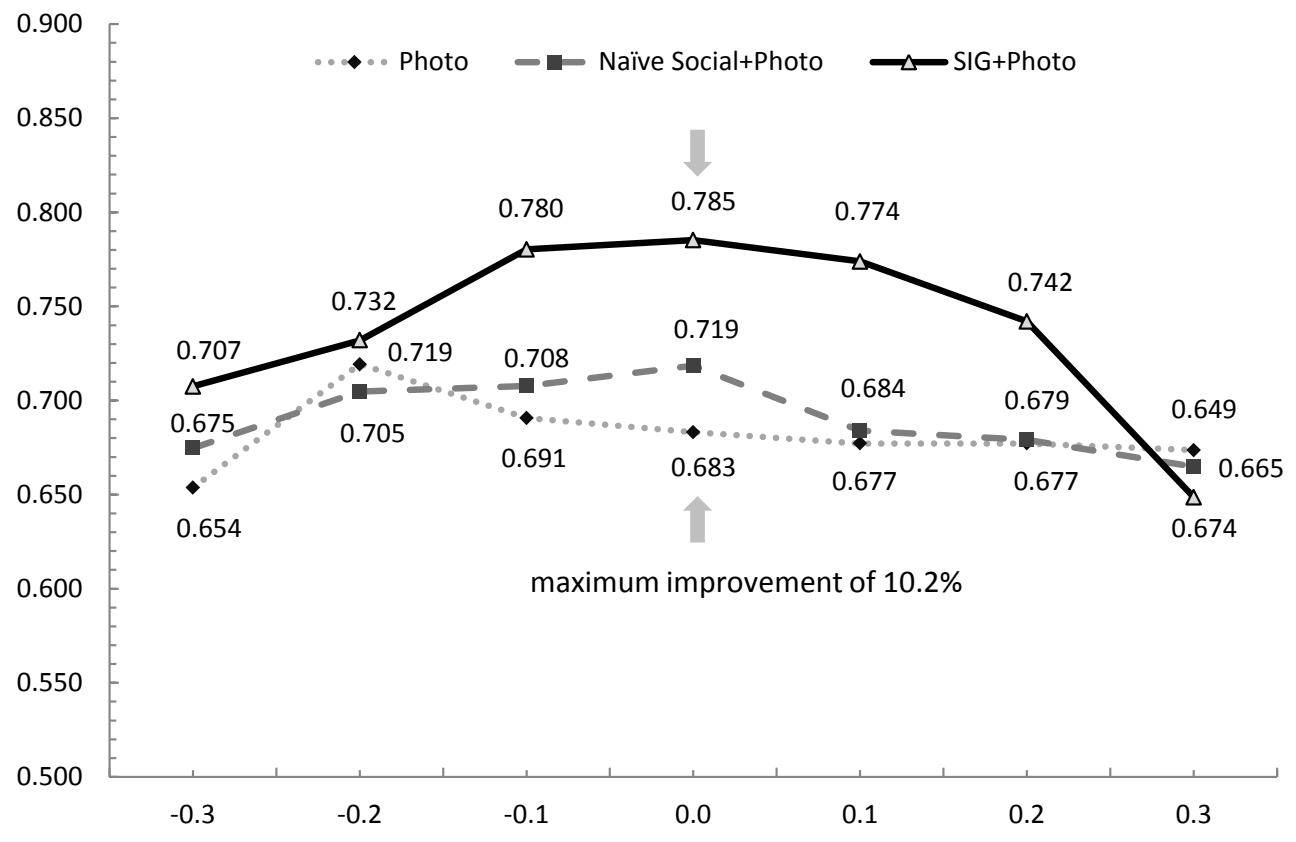

**Figure 2: F1 scores for the three methods on the Upcoming dataset. We show a maximum improvement of 10.2% over the baseline [1] (photo-features only).**

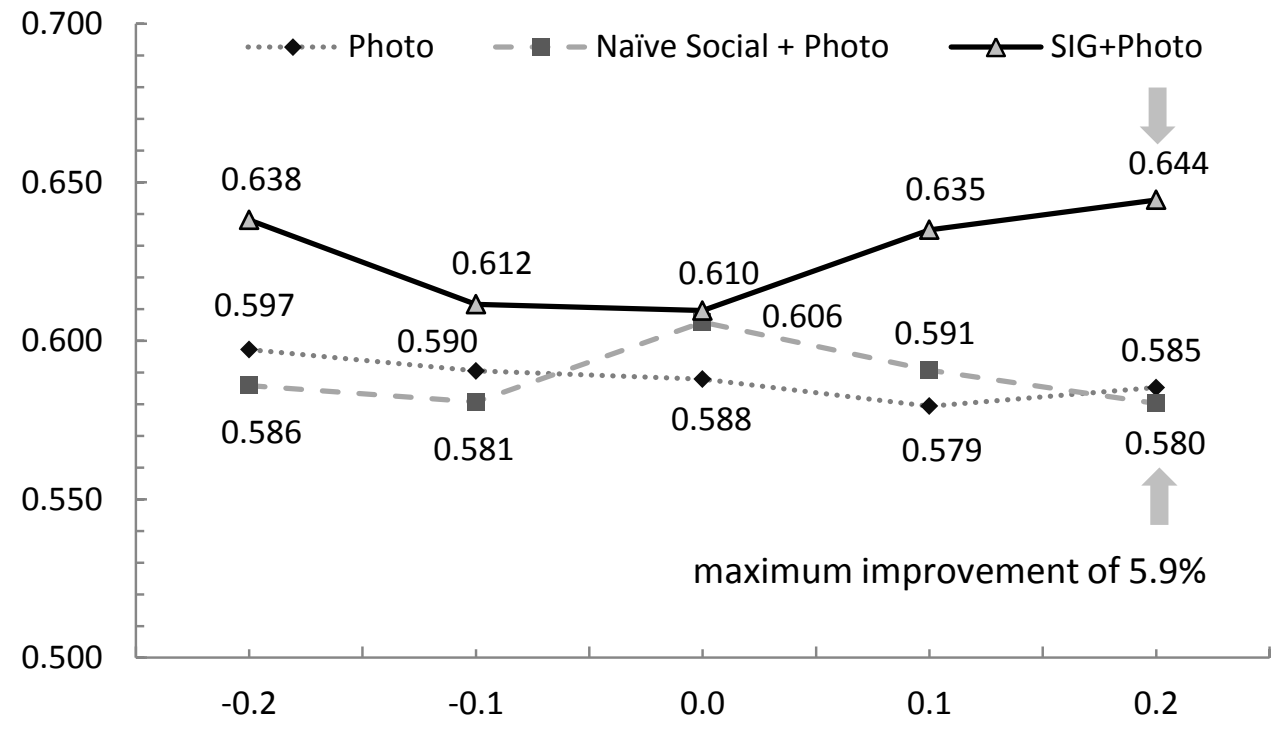

**Figure 3: F1 scores for the three methods on the MediaEval dataset. We show a maximum improvement of 5.9% over the baseline [1] (photo-features only).**

#465239 and #1280453, both group trips. The metadata feature on both photos(i.e. tags and text) does not overlap, hence making it a difficult task for the baseline approach which only uses photo features. There is no location data for either photo. The baseline approach considers them to be separate events. In contrast, our social interaction graph based approach takes advantage of social interactions by the multiple event participants on both photos, and successfully clusters them into the same event.

# 6. CONCLUSION

The goal of this paper was to detect social, physical-world events from photos posted on social media sites including Flickr. The main contribution of this paper was to incorporate social features in the detection of physical-events. We noted that prior work has *omitted social interaction as a feature* to detect events. We used a random-walk based approach to compute social affinity; this feature in addition to photo-centric metadata including time, location, tags and photo description to detect events. We used an SVM classifier in conjunction with an incremental clustering method to detect events. We tested our algorithm on two large datasets—MediaEval and Upcoming—with significant improvement from non-social baseline. Our goal was to demonstrate the utility of social interaction over baseline photo-features, despite that the final performance is not yet on par with the strongest in MediaEval 2011.

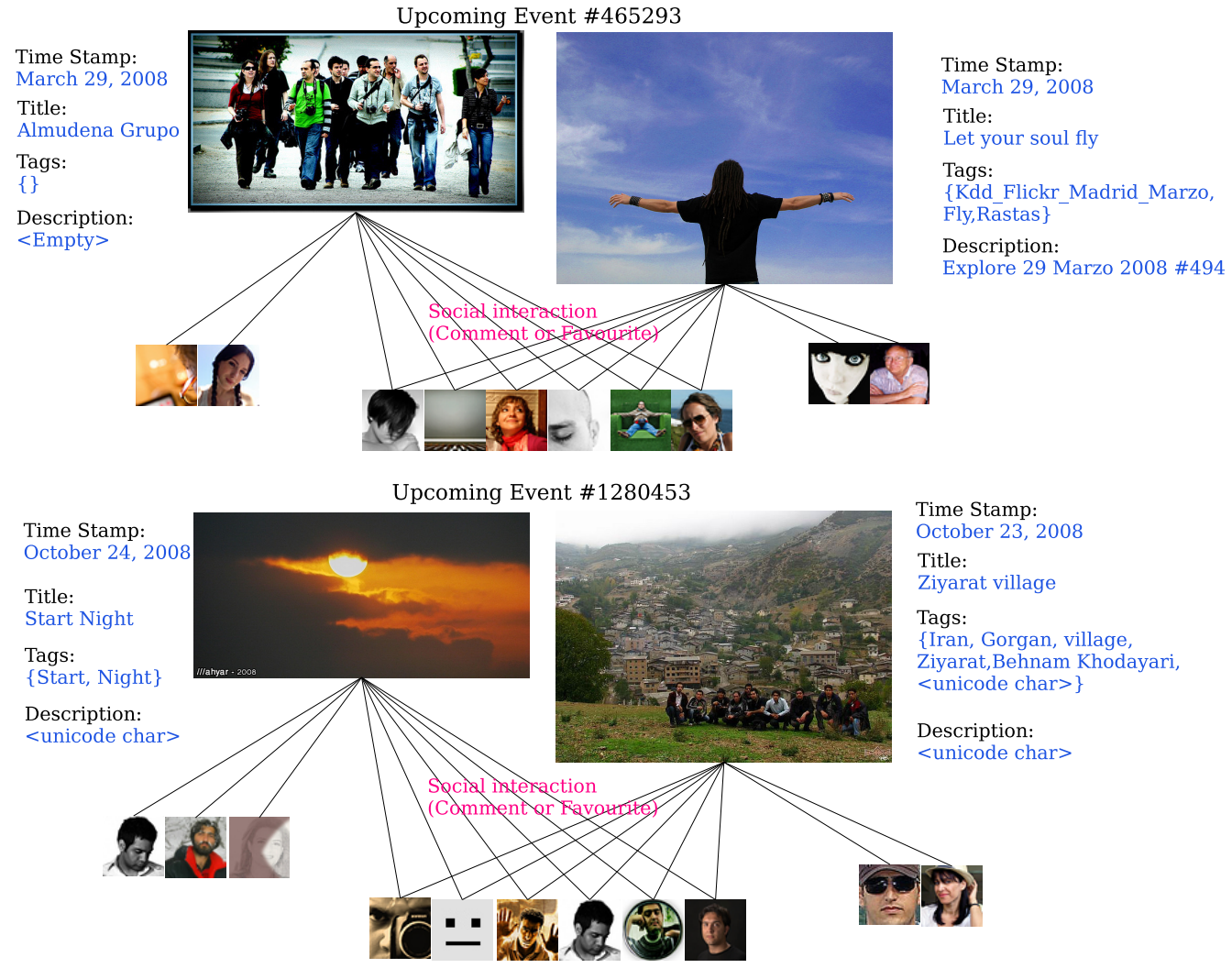

**Figure 4: Our proposed Social Interaction Graph approach can successfully group these two photos into the same event, whereas the baseline photo-features based method fails. Photos thumbnails courtesy of Flickr users Dakotilla, behn@m, ///ahyar, Pablo Kittsteiner.**

We plan to improve our results by further tuning image- and text-based filtering methods coupled with social interaction.

# Acknowledgments

NICTA is funded by the Australian Government as represented by the Department of Broadband, Communications and the Digital Economy and the Australian Research Council through the ICT Centre of Excellence program.